# Electric field enhances the electronic and diffusion properties of penta-graphene nanoribbons for application in lithium-ion batteries: a first-principles study

Thi Nhan Tran[1], Nguyen Vo Anh Duy[2], Nguyen Hoang Hieu[3], Truc Anh Nguyen[4], Nguyen To Van[5], Viet Bac Thi Phung[6,7], Peter Schall[8], and Minh Triet Dang*[3]

[1]Faculty of Fundamental Sciences, Hanoi University of Industry, 298 Cau Dien, Bac Tu Liem, Hanoi 100000, Vietnam.
[2]FPT University, 600 Nguyen Van Cu, Ninh Kieu, Can Tho, Vietnam.
[3]Can Tho University, 3/2 Street, Ninh Kieu, Can Tho, Vietnam.
[4]Faculty of Mechanics, Can Tho University of Technology, 256 Nguyen Van Cu Street, Ninh Kieu, Can Tho, Vietnam
[5]Faculty of Chemico-Physical Engineering, Le Quy Don Technical University, Ha Noi 100000, Vietnam
[6]Center for Environmental Intelligence and College of Engineering & Computer Science, VinUniversity, Hanoi 100000, Vietnam.
[7]Institute of Sustainability Science, VNU Vietnam Japan University, Vietnam National University, Hanoi 100000, Vietnam
[8]Van der Waals-Zeeman Institute, University of Amsterdam, Science Park 904, Amsterdam, The Netherlands
* Email: dmtriet@ctu.edu.vn

**Abstract.** Enhancing the electronic and diffusion properties of lithium-ion batteries is crucial for improving the performance of the fast-growing energy storage devices. Recently, fast-charging capability of commercial-like lithium-ion anodes with the least modification of the current manufactoring technology is of great interest. Here we use first principles methods with density functional theory and the climbing image-nudged elastic band method to evaluate the impact of an external electric field on the stability, electronic and diffusion properties of penta-graphene nanoribbons upon lithium adsorption. We show that by adsorbing a lithium atom, these semiconductor nanoribbons become metal with a formation energy of - 0.22 (eV). The lithium-ion mobility of this material is comparable to that of a common carbon graphite layer. Under a relatively small vertical electric field, the structural stability of these lithium-ion systems is even more stable, and their diffusion coefficient is enhanced significantly of ~719 times higher than that of the material in the absence of an applied electric field and ~521 times higher than in the case of commercial graphitic carbon layers. Our results highlight the role of an external electric field as a novel switch to improve the efficiency of lithium-ion batteries with penta-graphene nanoribbon electrodes and open a new horizon for the use of more environmentally friendly pentagonal materials as anode materials in lithium-ion battery industry.

## 1. Introduction

Lithium-ion rechargeable battery industry is a fast-growing market for mobile devices and electric vehicles [1]. Due to the rapid demand of the society and the shortage of raw material supply, the cost of rechargeable lithium-ion batteries (LIBs) is currently very high and keeps increasing [2]. This drawback limits the applicability of these types of batteries both in industry and for societal needs. The urgent need is to reduce the cost for fabricating LIBs and to improve the electrochemical performance such as high reversible capacity, high conductivity, reasonable cycling life and environmentally friendly manufacturing. To boost commercial mobile devices and electric vehicles, besides increasing the storage capacity of the batteries, fast-charging capability is urgently needed. Thus, reducing the energy barrier of lithium-ion diffusion is crucial to improve the electrochemical performance of these fast-growing energy storage devices.

Typical LIBs consist of anode and cathode electrodes, a separator, and all immersed electrolyte solution. To fulfil the aforementioned requirements, two critical tasks are improving the performance of cathode and/or anode electrode materials. Several commercial types of cathode electrode materials for LIBs are $LiCoO_2$ (LCO) [3], [4], $LiMnO_2$ (LMO) [5], [6], $LiFePO_4$ (LPO) [7], [8], $LiPF_6$ [9], [10] and $Li[Ni_xMn_yCo_z]O_2$ (NMC) [11]. For long lifetime and large energy storage LIBs, nickel-rich NMC layered cathodes can be considered as an excellent candidate to reconcile the requirement of high specific discharge capacity, reasonable durability and working voltages, and affordable production cost [12]. Recently, we proposed a modified co-precipitation method to synthesize $Li_{1.0}Ni_{0.6}Mn_{0.2}Co_{0.2}O_2$ materials, which overcomes the limitations of traditional material synthesis approaches [13]. The proposed method can be operated in a lower temperature environment with respect to the requirements of conventional synthesis methods. In the case of anode materials, the most common electrode material for commercial lithium-ion rechargeable batteries is graphite [14]. The advantages of graphite electrodes are low price, large reserves, and high conductivity. However, the disadvantage of the type of material is its low maximum storage capacity (372 mAhg$^{-1}$) due to the very flat surface and the fact that the charge/discharge rate tends to decrease rapidly after each charge/discharge cycle due to the formation of solid-electrolyte interface (SEI) [15]. Therefore, finding new electrode materials that possess the superior properties of graphite electrodes but can overcome their inherent disadvantages (storage capacity and electron diffusion ability of metal ions) is a major challenge in the lithium-ion rechargeable battery industry.

In 2015, Zhang *et. al.* theoretically proposed two-dimensional T12-carbon penta–graphene (PG) as a completely new type of hypothetical carbon material assembled from pentagonal bonds of carbon atoms [16]. These PG sheets have intrinsic indirect bandgap of 3.25 eV, predicted by the Heyd-Scuseria-Ernzerhof (HSE06) exchange-correlation functional, and are thermally stable at temperatures up to 1000K. In 2016, using density functional theory and molecular dynamics simulations, Xiao *et. al.* theoretically showed that albeit penta-graphene sheets are mechanically less stable than graphite monolayers, their maximum storage capacity and average open circuit voltage (OCV) and electron diffusion barrier are 1489 mAhg$^{-1}$, 0.55 V and 0.17 eV, respectively, making them outstanding candidates for lithium-ion battery anodes [17]. These nanoribbons feature high carrier mobility from ~$10^1$ to ~$10^4$ cm$^2$V$^{-1}$s$^{-1}$ at room temperature depending on the edge-terminated atoms [18]. With hydrogen termination at the edges, the PGNR behaves as a semiconductor with a bandgap of 2.4 eV [18] and a low energy diffusion barrier of 0.40 eV [19].

Recently, substantial progress has been made to meet the fast-charging requirement by applying external magnetic [20], [21] or electric fields [22]. Recent experimental reports demonstrate that the charge rates of LIBs can be significantly improved by imposing an external magnetic field parallel to the material surface to reduce the Li-ion transmission paths and increase the diffusion coefficients [21]. For instance, by applying an external electric field on lithium hexafluorophosphate (LiPF$_6$) in ethylene carbonate, Kumar and Seminario [9] showed a strong impact of the field strength on the electron migration rate of metal ions. Specifically, while the diffusion coefficient increased linearly under a small applied electric field with magnitude below 2 V/nm, it increased exponentially when the electric field strength was above that threshold [9]. Shi *et. al.* also showed theoretically that by applying electric fields on MoS$_2$ monolayers, upon adsorption of Li, Mg, and Al atoms, the electron diffusion barriers decrease while the charge rates of the batteries are accelerated effectively [22]. These results demonstrate the essential role of electric and magnetic fields for improving the migration speed of free electrons for faster and safer charging of LIBs.

Here we employ first principles methods with density functional theory and climbing image-nudged elastic band method to investigate the Li-ion adsorption, electronic properties, and diffusion for PGNR anodes in lithium-ion batteries. We demonstrate the greatly improved electrochemical properties of penta-graphene anodes resulting from the change in structural stability, density of states, electron diffusion and transition energy under applied electric fields. Such improvement demonstrates the potential application of penta-graphene nanoribbons as anode materials for the next generation of lithium-ion rechargeable batteries.

## 2. Computational methods

In this study, all periodic DFT optimizations are performed using the projector-augmented wave (PAW) method [23] implemented in the Vienna *Ab initio* simulation package (VASP) [24], [25]. We also include the optB86b-vdW [26] exchange-correlation functional in the vdW-DF family to provide reasonable comparison to experimental observations for graphene-like materials [27]. The integration in the Brillouin zone was employed using the Monkhorst-Pack scheme [28] (11 x 1 x 1) with an energy cut-off of 500 eV. The convergence threshold for the self-consistent field calculations was set to $10^{-5}$ eV per cell, and the geometrical structures were fully optimized until the Hellmann-Feynman forces acting on atoms were less than 0.01 eV Å$^{-1}$. To avoid the boundary effect on the interatomic layers of PGNR, a vacuum of 20 Å is applied between two neighboring nanoribbons.

To determine the most stable of PGNR with a single Li atom, we place the Li atom 3 Å above the top layer of the PGNR at all possible adsorption sites as shown by letters A-I in Fig. 1(a). The stability of the adsorption configurations is evaluated by the formation energy as follows [29]

$$E_{Form} = \frac{E_{total} - n_C E_C - n_H E_H - n_{Li} E_{Li}}{n_C + n_H + n_{Li}},$$

in which $E_{total}$ is the total Gibbs free energy of the adsorption systems, $E_C$, $E_H$, $E_{Li}$ stand for the chemical potentials of C, H, Li atoms. The symbols $n_C$, $n_H$, $n_{Li}$ are the number of C, H, Li atoms in the hosted PGNR samples.

To study the electronic properties of the pristine PGNR and the PGNR upon lithium adsorption at high accuracy, we apply the Heyd-Scuseria-Ernzerhof (HSE06) [30] hybrid functional implemented in the VASP (with the k-point grid of 7 x 1 x 1 in the Monkhorst-Pack scheme) and in the BAND engine of the Amsterdam Modeling Suite (AMS) [31] packages. For the band structure and density of state calculations in the BAND engine, we apply the triple zeta polarization (TZP) basic sets with the core orbitals considered as frozen during the self-consistent field procedure. For further analyses of orbital hybridization, we perform the periodic energy decomposition analyses (PEDA) using the BP86-D3 exchange-correlation potential with the same basic sets in the AMS package.

The charge/discharge capability is an essential factor to determine whether a proposed material is suitable for developing an anode material for lithium-ion batteries. To characterize the diffusion behavior of lithium ions, we use the climbing image-nudged elastic band (NEB) [32], [33] implemented in the AMS package. The initial and final states were taken systematically from the geometrical optimization procedure in VASP. For each migration path, sixteen intermediate images were used during the NEB image calculations.

## 3. Results and discussion

### 3.1 Adsorption of Li-ion on penta-graphene nanoribbons

To evaluate the effect of an external electric field on the electronic and diffusion properties of PGNR upon lithium adsorption, we perform DFT optimization on a PGNR with a width of nine sawtooth carbon chains to obtain the most stable pristine configuration as shown in Fig. 1(a). The C atoms are distributed in three hierarchical atomic planes with a roughness of $d \approx 0.768$ Å. The thickness of these nanoribbons (measured from the top to the bottom layers of C atoms) is $h = 1.412$ Å. There are two types of C-C bonds in the pristine PGNR: C-C single bonds with a length of ~ 1,504 - 1,569 Å and C=C double bonds with a length of ~ 1,337 - 1,341 Å. The $sp^2$ hybridized carbon atoms with three coordination numbers (denoted as $C_1$ atoms) are located in the middle layer, while the $sp^3$ carbon atoms with four coordination numbers ($C_2$) lie in the bottom and top layers (Fig. 1(a)-(c)). These results agree well with previous reports [16], [34], [35]. To further evaluate the stability of PGNR, we also calculate its phonon dispersion spectra. As shown in the phonon spectra of Fig. S1(a), these ribbons are dynamically stable due to the absence of imaginary modes, and we also observe distinct acoustic modes close to the Γ-point, illustrating the typical out-of-plane, in-plane longitudinal, and in-plane transverse atomic motions typical for (penta-)graphene, as thoroughly discussed in Ref. [16].

To investigate the adsorption of Li atoms, we evaluate the structural stability upon adsorption of a single Li atom on nine possible adsorption sites of the most stable PGNR configuration (Fig. 1a). Table 1 presents the formation energy $E_{ads}$ and the corresponding adsorption distances of all possible adsorption sites. The adsorption distance is the center-to-center distance from the adsorbed Li atom to the nearest C atoms of the PGNR. As shown in Table 1, the lowest adsorption energy of all possible adsorption sites is -0.220 eV at the D and E sites. Furthermore, after equilibration, regardless of the initial configurations at D or E sites, the adsorbed Li atom relaxes on top of the $sp^3$ hybridized C atom, illustrating that the $sp^3$ configuration is the most energetical stable one. The fundamental phonon dispersion characteristic remains almost unchanged for these nanoribbons upon lithium adsorption (Fig. S1(b)). Thus, for the further calculations, we associate the PGNR with a Li atom on top of the $sp^3$ C atom as the most stable configuration.

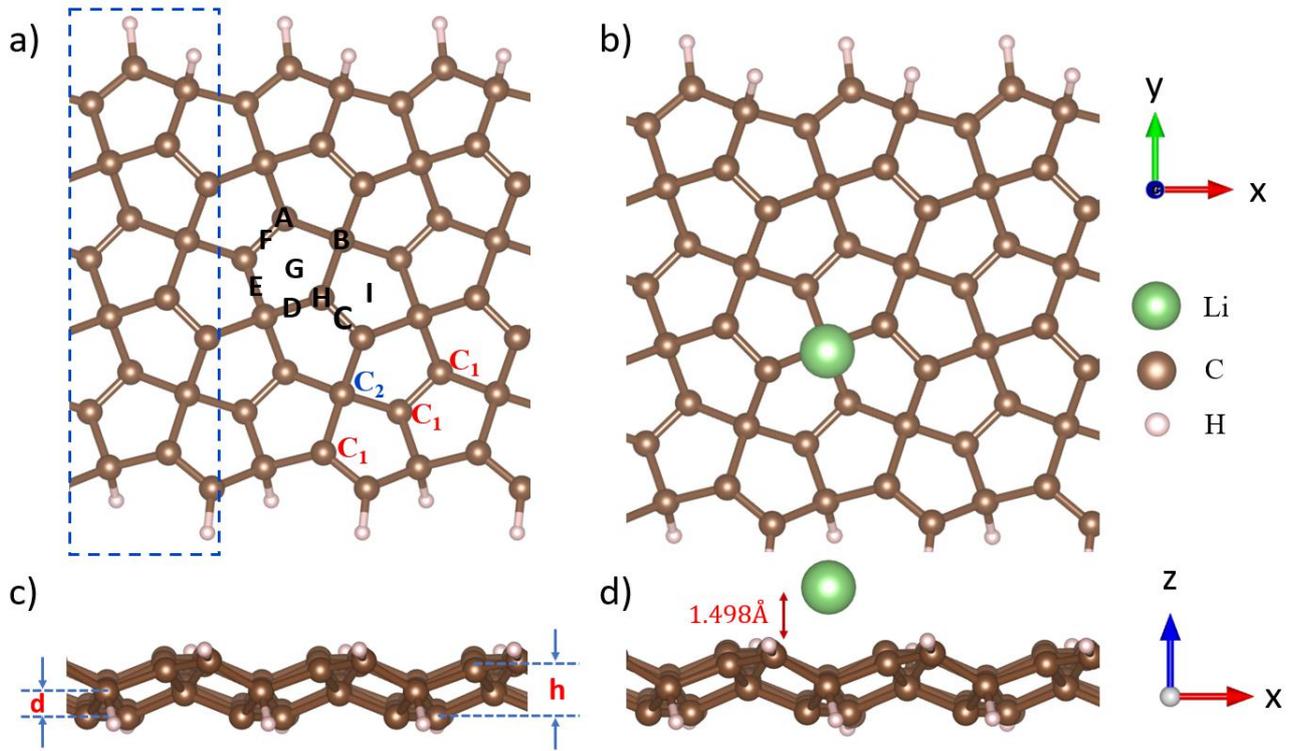

**Figure 1.** Top- and side-view of the optimized geometrical structures of PGNR (ac) and PGNR + Li (bd). The yellow, gray and white balls illustrate Li, C and H atoms, respectively. The possible Li adsorption sites (a) are marked as on top (A, B, H), at hollow (I, G), and in the bridge (C, D, E, F) of C atoms. The dashed blue rectangle represents the supercell of PGNR.

**Table 1.** Formation energies $E_{ads}$ and adsorption distances $a$ from the Li atom to PGNR at different adsorption sites.

| Site | A | B | C | D | E | F | G | H | I |
|---|---|---|---|---|---|---|---|---|---|
| $E_{ads}$ (eV) | -0.219 | -0.219 | -0.216 | -0.220 | -0.220 | -0.219 | -0.218 | -0.218 | -0.219 |
| $a$ (Å) | 1.345 | 1.444 | 1.943 | 1.498 | 1.491 | 1.014 | 1.026 | 1.031 | 1.148 |

**Table 2.** The structural modification of the PGNR under an applied electric field.

| Electric field strength (V/nm) | -2 | -1 | 0 | 1 | 2 |
|---|---|---|---|---|---|
| Relative energy difference (eV) | -0.018 | -0.162 | 0 | -0.084 | -0.169 |
| Lattice constant in x-direction (Å) | 10.854 | 10.854 | 10.857 | 10.854 | 10.855 |
| Buckling distance (Å) | 0.642 | 0.743 | 0.677 | 0.712 | 0.667 |

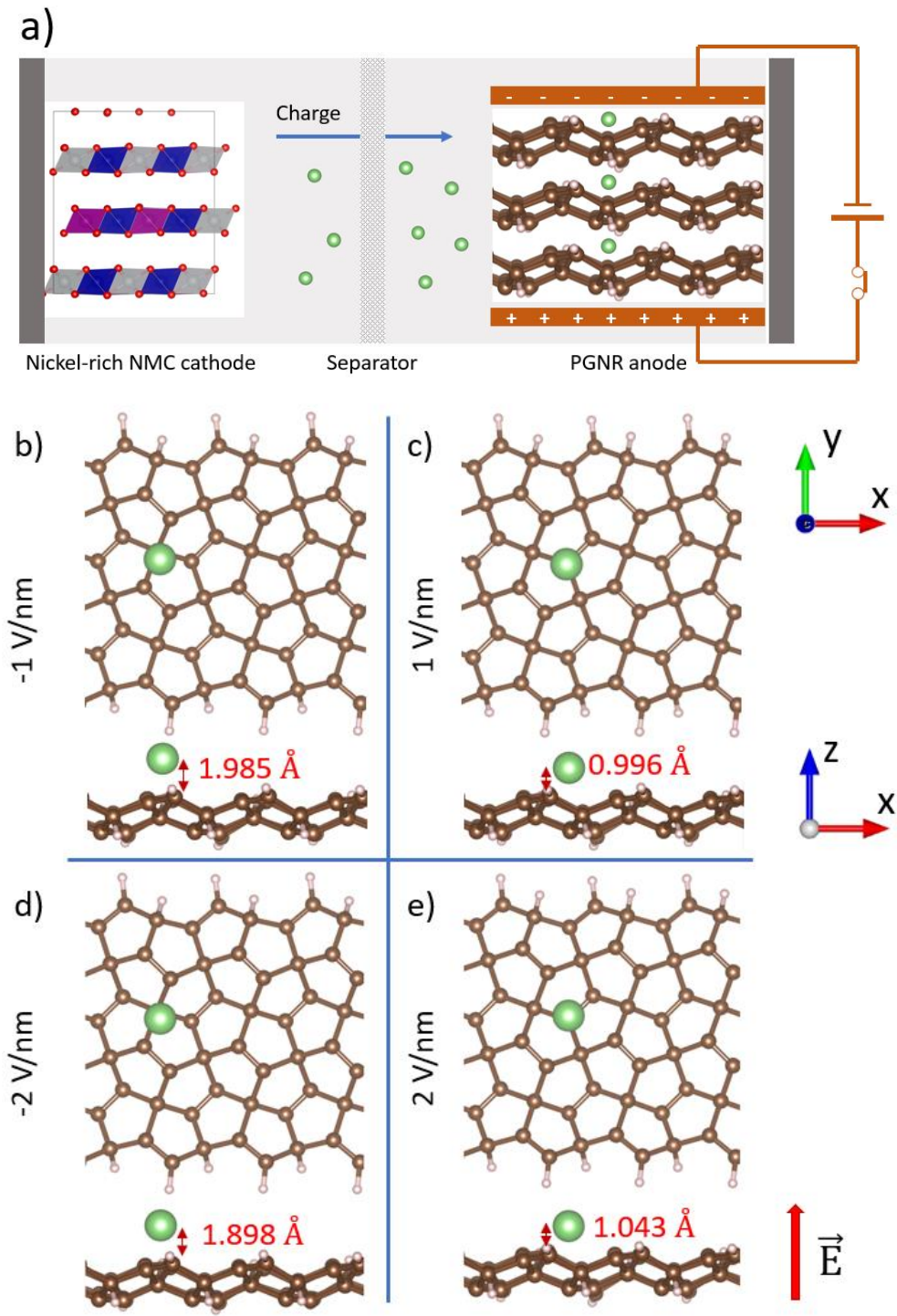

**Figure 2.** (a) Sketch of a lithium-ion battery with a penta-graphene nanoribbon anode under a vertical electric field. The geometrical structures with top and side views of these adsorbed system under an electric field of -1V/nm (b), 1 V/nm (c), -2 V/nm (d) and 2 V/nm (e). The positive direction of the external electric field is sketched at the bottom right of figure (e). The shortest distances from the adsorbed lithium-ion to PGNR are also displayed in these figures.

We now apply an electric field perpendicular to these nanoribbons. The strength of the applied field varies from -2 V/nm to 2 V/nm with a stepsize of 1 V/nm, where positive (negative) values indicate a field in the positive (negative) z-direction. This field is strong enough to disturb the electron diffusion process but is weak enough not to destroy the adsorption system. Under a positive electric field, the adsorbed Li atom relaxes on top of the $sp^2$ hybridized carbon atom as indicated in Fig. 2. However, under a negative electric field, the adsorbed Li atom anchors close to the top of the $sp^3$ hybridized carbon atom. To evaluate the stability of these configurations, we calculate the relative energy difference, which is the energy difference between the systems with the electric field and the one without. As shown in Table 2, the adsorption systems are slightly more stable with applied electric field than without, as indicated by the negative values of the relative energy difference. Apart from that, the lattice constant in the x-direction and the buckling amplitude of the systems show minor changes with the largest buckling of 9.75%. Though there is a presence of a negative dispersion band when these nanoribbons are under a positive electric field (Fig. S1(c-d)), in general, these ribbons are mechanically stable under these small electric fields. The most interesting observation is that the Li atom is adsorbed more closely to the PGNR when a positive electric field is applied. As shown in Fig. 2, the adsorption distance from the Li atom to the PGNR reduces by about 30% when the field is applied in the z-direction, while for fields in the opposite direction, this quantity extends by about 30% with respect to the field-less case. This implies significant changes of the electronic transport properties as discussed in the next section.

**3.2 Effects of an electric field on the electronic properties of Li-ion penta-graphene nanoribons**

To increase the accuracy as well as the reliability of the calculated electronic properties, we further calculate the band structure and partial density of states of these adsorbed systems using the state-of-art HSE06 hybrid functionals [36]. As shown in Figs. 3 and S2-S3, the PGNR exhibits semiconductor properties with an indirect bandgap of 3.415 eV or 3.433 eV, as calculated by the AMS and VASP packages, respectively. These results are in quantitative agreement with the previous report of 3.25 eV [16] for a two-dimensional penta-graphene system. The electronic energy states are mainly dominated by the 2-$p_z$ orbitals of carbon atoms (Fig. S3), illustrating the high possibility for orbital hybrization in the $z$ direction. The localized $\pi$ electrons of $sp^2$ carbon atoms are also well visualized by the shapes of frontier orbitals (both the highest occupied molecular orbital (HOMO) and the lowest unoccupied molecular orbital (LUMO) states) on the top-left of Table 3, which is typical for penta-graphene nanoribbons [37].

Upon lithium adsorption, the semiconductor PGNR becomes metal as indicated in Fig. 4, in line with previous reports [16], [38]. This semiconductor-to-metal transition demonstrates that PGNR is a good anode material for lithium-ion batteries due to creation of free charges from the interaction between the adsorbed Li atoms and C atoms of the PGNR. The energy states around the Fermi level are driven by the π electrons of carbon atoms as displayed in Fig. S4, demonstrating a charge transfer from the adsorbed Li atom to the PGNR as pointed out in Ref. [38] and a high possibility of hybridization between *p*-orbital of the C and *s*-orbital of the Li atom. The significant contribution of the Li-atom's *s*-orbital to the density of states close to the Fermi level cause an increase in the conductivity of PGNR upon lithium adsorption and lead to the semiconductor-to-metal transition. Additionally, the interaction with the Li atom can enhance the flexibility of *p*-orbital electrons of the C atoms, leading to a strong shift of the highest occupied molecular orbital (HOMO) towards to the bottom of the conduction band (see Figs. 4 and S4). The electrons transferred from the Li atom, and the flexible *p*-orbital electrons of the C atoms occupying the HOMO play an important role in increasing the conductivity of the PGNR + Li systems.

A significant band modification of the PGNR + Li under an external electric field is clearly visible in Fig. 4. Under a 2 V/nm electric field, we observe a considerable down-shift of the band dispersion curves in the conduction band towards the Fermi level while this trend is less pronounced in the case of a – 2 V/nm electric field. Similar behavior is also observed in Fig. S4 obtained from the VASP package. Furthermore, under a positive applied electric field, a new state appears in the conduction band just above the Fermi level. This state is occupied by *s*-orbitals of the adsorbed Li atoms, increasing its pDOS compared to nearby states (Fig. S4) further increasing the conductivity of PGNR + Li under positive electric fields. We stress that these modifications of the band structure and pDOS are different from those under negative electric field; hence, the applied field direction has a strong impact on the electronic characteristics of the PGNR + Li systems.

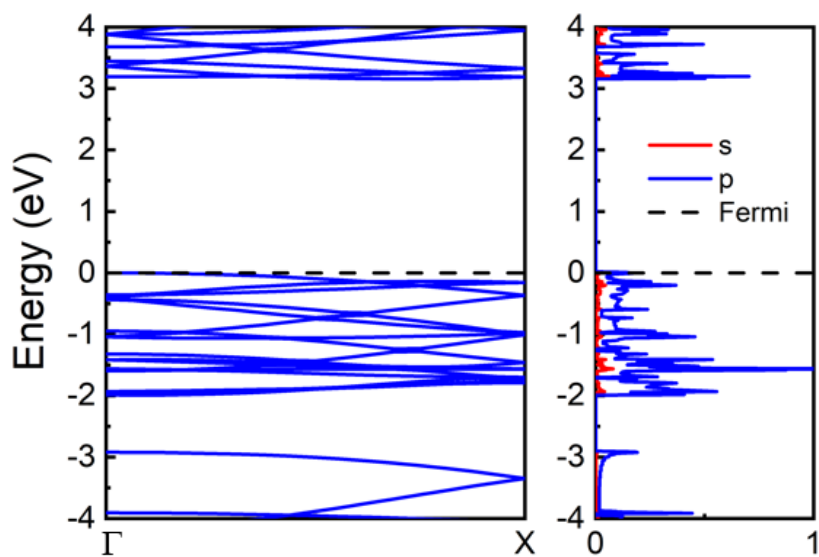

**Figure 3.** Band structure and partial density of states of penta-graphene nanoribbons calculated in the AMS package. The Fermi level is indicated by the dashed lines.

**Table 3**. Shapes of frontier orbitals of PGNR and PGNR upon lithium adsorption in the absence and under an external electric field in top (upper images) and side (lower images) views. These shapes are calculated in the AMS package. Iso-value for the orbital rendering is 0.02 eV/Å$^2$. The electron depletion is in blue whereas the electron accumulation is in reddish brown.

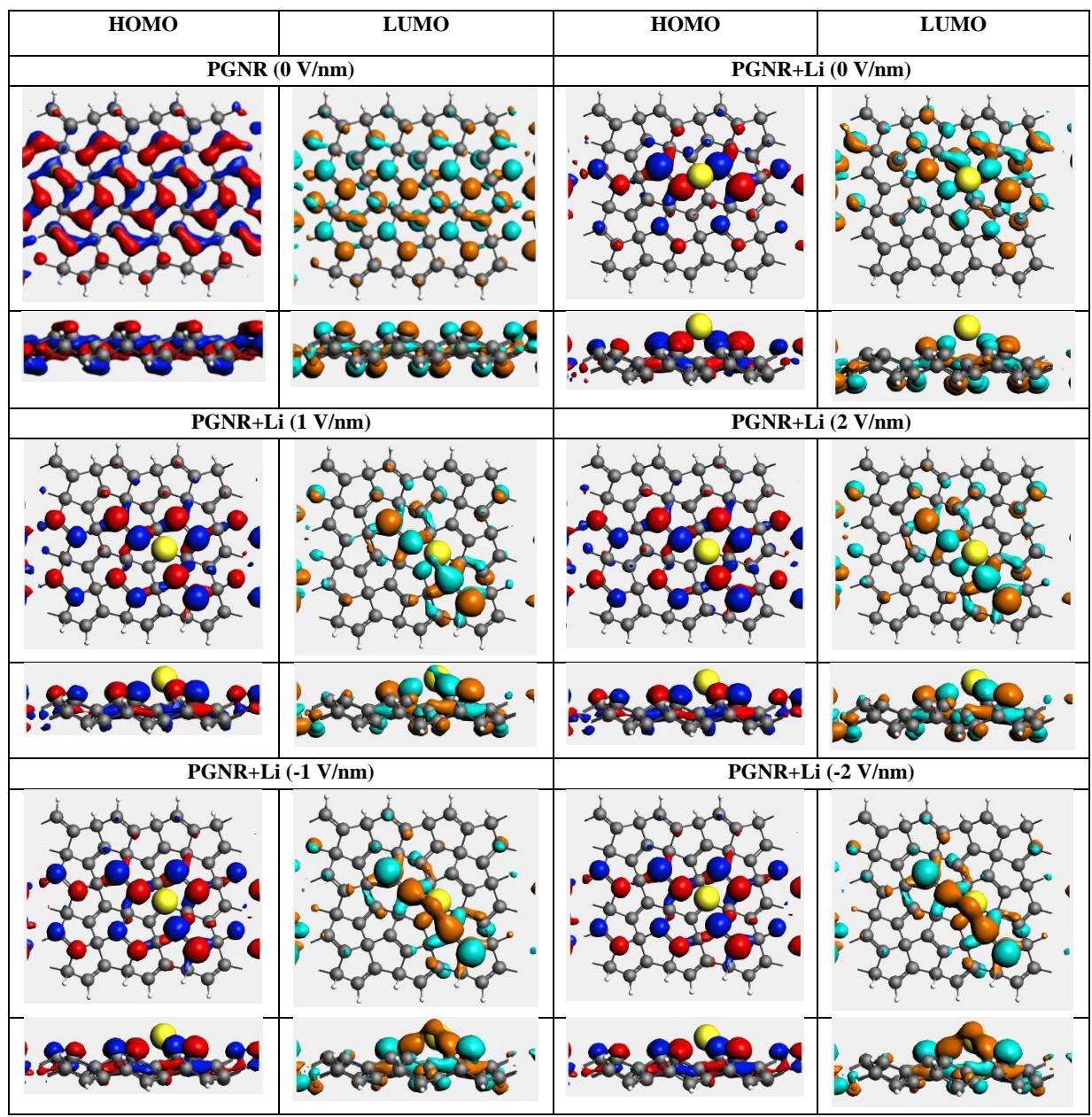

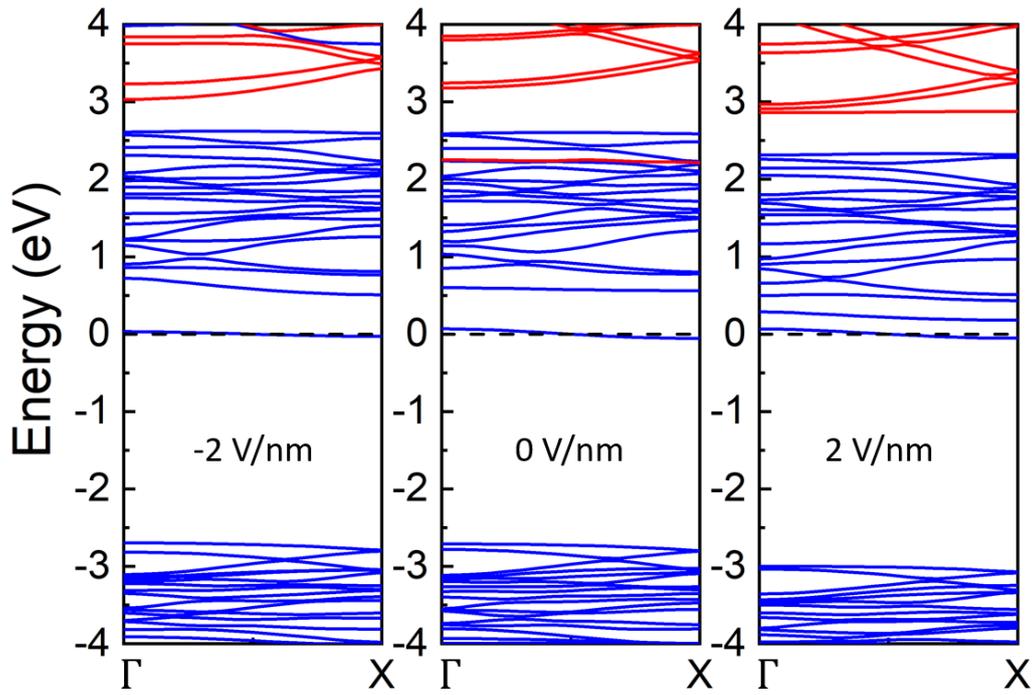

**Figure 4.** Band structure of penta-graphene nanoribbons upon lithium adsorption calculated in the AMS package. Red and blue lines represent for the s- and p-orbitals. The Fermi level is indicated by the dashed lines.

The electrons of the frontier orbitals play an important role, governing the electronic properties of the material. To further insight the change in electronic properties of PGNR caused from the Li adsorption and the effect of the external electric field, we probe the modification of the local electron distribution of HOMO and LUMO states of the material. Table 3 displays the shapes of frontier orbitals of pristine PGNR, PGNR + Li without and with an external electric field. For pristine PGNR (without applied external electric field), electrons of the HOMO and LUMO mainly localize around $sp^2$ C atoms and distribute over the areas inside the PGNR like that observed for PG [16]. The density of electrons of these two states around C atoms at the edge of the nanoribbon is inconsiderable, possibly, due to effect of the H passivation. Upon lithium adsorption, the distribution of electrons of the HOMO and LUMO exhibits a higher localization. They mainly concentrate on the top of C atoms near the adsorbed Li atom with relative high density and could act as free electrons, leading to the increase of the conductivity of PGNR upon adsorption. Such a redistribution of electrons of the frontier orbitals due to the Li adsorption also results the transition from the conductor-to-metal behavior of PGNR. As shown in Table 3, under an electric field, the electrons at the LUMO states distribute more locally around the adsorbed Li atoms. Such change of the LUMO orbital shapes implies a strong increase of the conductivity of PGNR + Li under the external electric field. Interestingly, Table 3 also shows that under a negative electric field, the electron depletion areas of the frontier orbitals are replaced by the accumulation areas, whereas this phenomenon is not observed as applying a positive field, demonstrating the dependence of electronic properties of the PGNR + Li systems on the direction of the applied fields.

To further understand the influence of external electric fields on the electronic properties of the adsorbed systems, we perform periodic energy decomposition analyses (PEDA) to calculate the intrinsic bond energy $\Delta E_{int}$ consisting of the electrostatic energy $\Delta E_{elstat}$, the Pauli repulsion energy $\Delta E_{Pauli}$, and the orbital relaxation energy according to

$$\Delta E_{int} = \Delta E_{eslast} + \Delta E_{Pauli} + \Delta E_{orb}, \quad (7)$$

the computed intrinsic bond and partial energies are given in Table 4. These indicate that a weak external electric field can modify the intrinsic bond energies and their parts, by modifying the interaction between PGNR and Li atoms. Looking in more detail, the electrostatic energy is the most dominant factor in the absence or presence of external electric fields, illustrating that the interaction between PGNR and Li is mainly driven by electrostatic effects and the adsorbed Li atoms yield considerable electronic contributions to the PGNR. Looking at the orbital relaxation energies, which cover the orbital hybridization effect in the adsorption system, with respect to the zero electric field, the magnitudes of this quantity decrease under a negative electric field and increase significantly under the positive field, again indicating the sensitivity of these adsorbed systems under an external electric field. The increase in electrostatic energy under positive electric field results from a charge redistribution of the frontier orbital and the nearby states as shown in Table 3 and Figs. S4. Hence, the enhanced charge distribution of the adsorbent and adsorbate originates from charge transfer due to the reduced distance between the Li and PGNR as pointed out above. As a result, the diffusion coefficient of Li ions on PGNR under positive electric field is significantly higher than that without and under negative field. On the other hand, the smaller orbital relaxation energy under negative electric fields, implies a weakening of the orbital hybridization between the adsorbed Li and the PGNR. Therefore, positive electric fields enhance the flexibility of the *p*-orbital electrons of C atoms; and these results anticipate an increase of the conductivity as well as the diffusion coefficient of Li ions on the adsorbed systems.

**Table 4.** Periodic energy decomposition analyses (PEDA) of PGNR upon lithium adsorption calculated in the AMS package. All units are in kJ mol$^{-1}$.

| PEDA types | E = - 2 V/nm | E = 0 V/nm | E = 2 V/nm |
|---|---|---|---|
| $\Delta E_{int}$ | -41.9 | -64.9 | -62.2 |
| $\Delta E_{elstat}$ | -318.5 | -418.3 | -670.2 |
| $\Delta E_{orb}$ | -183.0 | -249.4 | -367.6 |

### 3.3 Effects of an electric field on the diffusion process of Li-ion penta-graphene nanoribbons

The diffusion energy profiles, and the energy barriers of lithium-ion migration are an essential factor for evaluating the performance of lithium-ion anodes. The low energy barriers indicate that the lithium-ions can easily migrate along preferable diffusion paths. In this section, we investigate the effect of an external electric field on the lithium-ion diffusion characteristics of penta-graphene nanoribbons. As discussed above, under a zero and negative electric fields, the adsorbed lithium atoms are energetically stable on top of the *sp*$^3$ hybridized carbon atoms while under the positive fields, the equivalent stabilized positions are on top of the *sp*$^2$ carbon atoms. The diffusion paths are chosen as the shortest paths between the two most energetically favorable configurations as illustrated by red arrows in Fig. 5. In the absence of an electric field, the diffusion barrier calculated by the CI-NEB method along the diffusion path highlighted in Fig. 5(a) is 0.272 eV, which is in quantitative agreement with that of two-dimensional penta-graphene ($\leq 0.33\ eV$) [17], comparable to the case of penta-hexagonal graphene (in between 0.21 eV and 0.32 eV) [39] and significantly lower than those of commercial graphite anodes (0.4-0.6 eV) [39] and LiFePO$_4$ (1.02 eV) electrodes [40]. In the presence of an external electric field, this diffusion barrier significantly reduces to the lowest value of ~0.1 eV at 2 V/nm electric field as shown in Table 5 and Fig. 5. Such low diffusion barrier, which is comparable with that of Ti$_3$C$_2$ [39] and the recent proposed MoS$_2$/penta-graphene heterostructures [41], strongly demonstrates the remarkable fast ion diffusion of penta-graphene nanoribbons under an applied electric field.

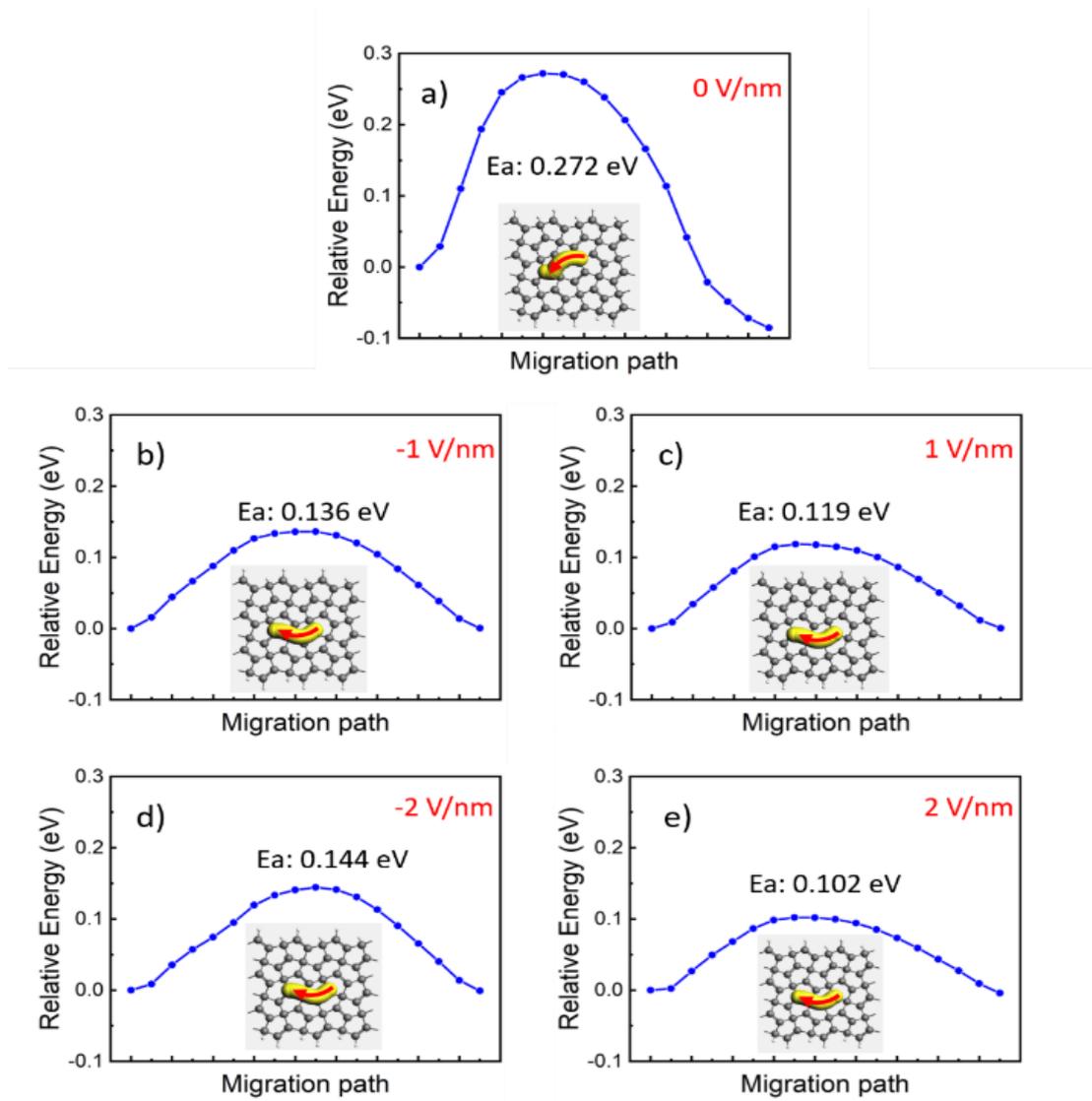

**Figure 5.** Diffusion barriers of lithium-atoms on penta-graphene nanoribbons under an external electric field.

The diffusion coefficient of lithium-ions can be calculated via the relation $D = a^2 v e^{-\frac{E_{act}}{k_B T}}$, where $a$ is the diffusion length of ions, $v \sim 10^{13} Hz$, $k_B$ is Boltzmann's constant, $T$ is the room temperature (T = 300K) [40], and $E_{act}$ stands for the diffusion barrier. As shown in Table 5, in the absence of an electric field, the lithium-ion diffusion coefficient of penta-graphene nanoribbons is $3.2 \times 10^{-6}$ cm$^2$/s while under an external electric field, this diffusion coefficient increases significantly. Specifically, under a positive electric field of 1 V/nm and 2 V/nm, it reaches to $1.2 \times 10^{-3}$ cm$^2$/s and $2.3 \times 10^{-3}$ cm$^2$/s, which is approximately about ~375 and ~719 times faster than that of the case of zero electric field, respectively. Under negative electric fields, though we also observe an increase of diffusion coefficients, these quantities are relatively smaller than in the case of positive electric fields. These observations are in line with the additional electronic states formed in the conduction bands close to the Fermi level (Fig. 4). To demonstrate the outperformance of PGNR as a promising material for lithium-ion anodes, we compare our calculated diffusion coefficients with those of carbon graphite layers, a common electrode material of commercial batteries [42]. Table 5 illustrates that the diffusion coefficient of PGNR under a zero electric field is slightly lower that of graphitic carbon layers but reaches ~521 times higher under a 2 V/nm electric field. Therefore, the external electric field can be used as a novel switch to remarkably improve the charge/discharge rate of penta-graphene lithium-ion anodes.

**Table 5**. Diffusion barrier and diffusion coefficient of lithium-ions on penta-graphene nanoribbons under an external electric field.

| Electric field strength (V/m) | -2 | -1 | 0 | 1 | 2 |
|---|---|---|---|---|---|
| Diffusion barrier (eV) | 0.144 | 0.13617 | 0.272 | 0.119 | 0.102 |
| Diffusion coefficient (cm²/s) | $4.5 \times 10^{-4}$ | $6.2 \times 10^{-4}$ | $3.2 \times 10^{-6}$ | $1.2 \times 10^{-3}$ | $2.3 \times 10^{-3}$ |
| The lithium diffusion coefficient ratio between penta-graphene nanoribbons and graphitic carbon [42] | 102.78 | 139.99 | 0.73 | 269.98 | 520.71 |

Besides the diffusion barriers and diffusion coefficients, another essential factor which greatly contributes to the charge/discharge rates of lithium-ion anodes is the electronic conductivity, which can be inferred by carrier effective mass [43]. The effective masses of electrons ($m_e^*$) and holes ($m_h^*$) are calculated from $m^* = \hbar^2 \left(\frac{d^2 E(k)}{dk^2}\right)^{-1}$ equation, where $\hbar$ is the Planck's constant, $k$ and $E(k)$ are respectively the wave vector and the energy dispersion relation corresponding to the conduction-band minima and valence-band maxima [40] of Figs. 3 and 4. The effective masses of holes and electrons of pristine PGNR are $0.852 m_0$ and $0.519 m_0$ (where $m_0$ is the mass of an electron), respectively, which are larger than the effective masses of holes and electrons in two-dimensional penta-graphene ($0.50 m_0$ and $0.24 m_0$, respectively [44]). These results indicate that holes and electrons of pristine penta-graphene nanoribbons are less mobile than those of penta-graphene monolayers. Upon lithium adsorption, the effective mass of electrons is $1.442 m_0$ which is ~2.8 times higher than that of pristine PGNR and the $m_h^*/m_e^*$ is 0.584, meaning that these electrons are even less mobile than in the case of pristine systems. However, under an external electric field, these $m_h^*/m_e^*$ effective mass ratios increase dramatically. While in the negative electric fields, these ratios increase a few times, in the opposite electric field directions, we observe remarkable increases up to ~26 times higher (see Table 6). Obviously, this sudden increase of the $m_h^*/m_e^*$ ratios highlight the role of an external electric field as a novel switch to improve the conductivity of penta-graphene lithium-ion anodes.

**Table 6**. Effective masses of electrons ($m_e^*/m_0$) and holes ($m_h^*/m_0$) of penta-graphene nanoribbons upon adsorption of lithium-ions under an external electric field.

| Electric field (V/nm) | -2 | -1 | 0 | 1 | 2 |
|---|---|---|---|---|---|
| $m_h^*/m_0$ | 0.865 | 0.863 | 0.842 | 0.864 | 0.865 |
| $m_e^*/m_0$ | 0.217 | 0.145 | 1.442 | 0.036 | 0.033 |
| $m_h^*/m_e^*$ | 3.986 | 5.952 | 0.584 | 24.000 | 26.212 |

## 4. Conclusions

We have presented in detail the remarkable role of an external electric field to enhance the electronic properties and the diffusion of lithium-ions of penta-graphene nanoribbons upon adsorption of lithium atoms. We show that the adsorbed systems are as thermodynamically stable as the pristine systems. After adsorbing lithium atoms, the semiconductor penta-graphene nanoribbons become metal, originating from the orbital hybridization between the *sp³* carbon atoms and the lithium atoms. PGNR anode materials exhibit fast diffusion coefficient of ~521 times higher than that of carbon graphite layer. The external electric field can be used a novel switch to improve the efficiency of lithium-ion batteries with penta-graphene nanoribbon anodes. Furthermore, these electric fields support directional orientations of electron conductivity and diffusions, and penta-graphene nanoribbons promise an excellent anode material for lithium-ion batteries with ultrafast charge/discharge rates.

## Conflicts of interest

There are no conflicts to declare.

## Acknowledgements

This research is funded by the Ministry of Education and Training, Vietnam (under Grant number B2024-TCT-16) and VinUniversity Seed Funding Program 23-24. The authors acknowledge the Information and Network Management Centre at Can Tho University for providing high-performance computing resources.

**Graphical abstract**

First principles calculations show a significant enhancement of the conductivity and the Li-ion diffusion coefficient of penta-graphene nanoribbons under external electric fields for application in fast-charging lithium-ion batteries.

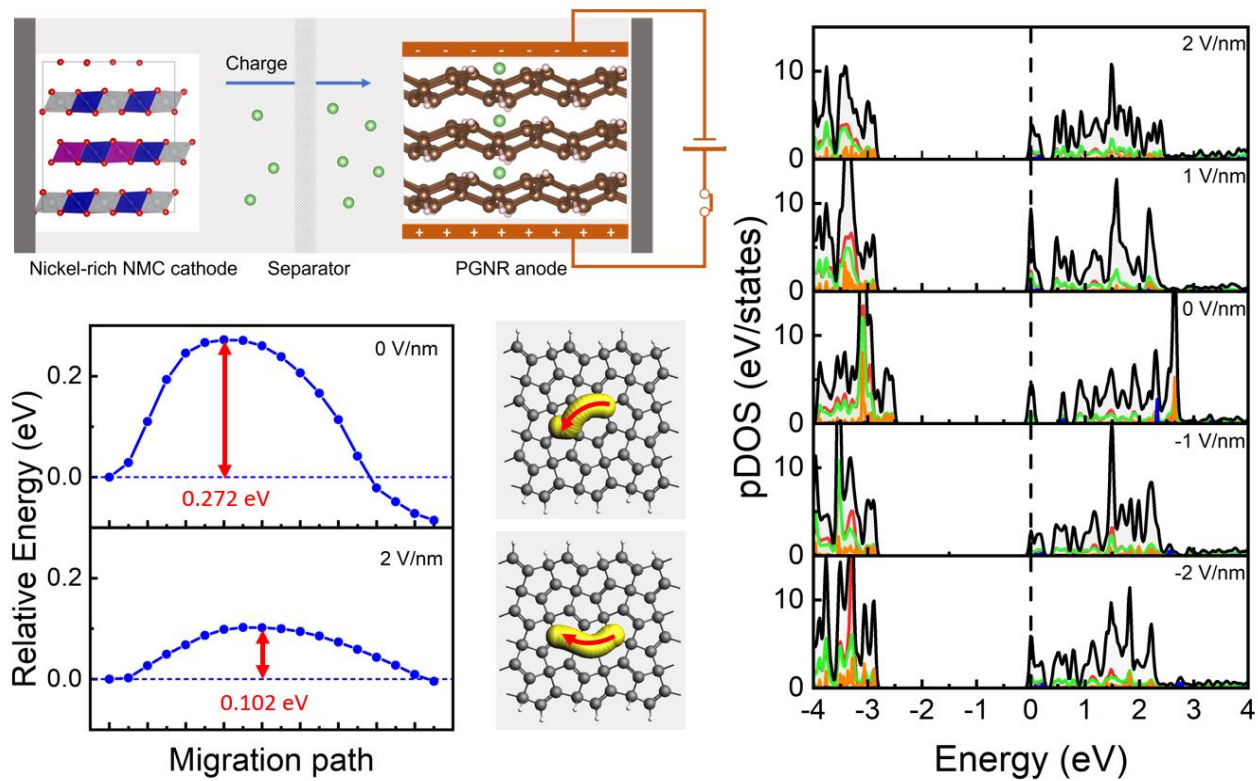